\documentclass[aps,prl,preprint,groupedaddress]{revtex4-1}
\usepackage{graphicx}

\begin{document}


\title{Revisiting the constraints on annihilating dark matter by radio observational data of M31}


\author{Man Ho Chan}
\affiliation{The Education University of Hong Kong}


\date{\today}

\begin{abstract}
Recent gamma-ray observations and radio observations put strong constraints on the parameters of dark matter annihilation. In this article, we derive new constraints for six standard model annihilation channels by using the recent radio data of M31 galaxy. The new constraints are generally tighter than the constraints obtained from 6 years of Fermi Large Area Telescope gamma-ray observations of the Milky Way dwarf spheroidal satellite galaxies. The conservative lower limits of dark matter mass annihilating via $b\bar{b}$, $\mu^+\mu^-$ and $\tau^+\tau^-$ channels are 90 GeV, 90 GeV and 80 GeV respectively with the canonical thermal relic cross section and the Burkert profile being the dark matter density profile. Hence, our results do not favor the most popular models of the dark matter interpretation of the Milky Way GeV gamma-ray excess.
\end{abstract}


\maketitle


\section{Introduction}
In the past few years, gamma-ray observations revealed the possibility of dark matter annihilation in our galaxy. Several groups claim that the excess GeV gamma rays emitted from our Galactic center can be best explained by dark matter annihilation through $b\bar{b}$ channel \cite{Daylan,Abazajian,Calore,Calore2}. The rest mass of dark matter and the required annihilation cross section are $m=30-50$ GeV and $<\sigma v>=(1.4-7.5)\times 10^{-26}$ cm$^3$ s$^{-1}$ respectively \cite{Daylan,Abazajian,Calore2}. However, recent studies of the Milky Way dwarf spheroidal satellite galaxies (MW dSphs) challenge the above claim. The constrained annihilation cross sections lie below the canonical thermal relic cross section ($<\sigma v> \approx 2.2\times 10^{-26}$ cm$^3$ s$^{-1}$) for dark matter of mass $\le 100$ GeV annihilating via quark and $\tau$-lepton channels \cite{Steigman,Ackermann}. As a result, only a very small parameter space remains possible for the dark matter interpretation of the GeV gamma-ray excess. Therefore, it is very important for us to search another independent observation to verify the claim.

Besides gamma-ray observations, radio observation is another way to test the dark matter annihilation model. It is commonly believed that many high-energy electron-positron pairs are produced from dark matter annihilation. These high-energy electrons and positrons would emit strong synchrotron radiation when there is a strong magnetic field. In fact, radio observations can give a stringent constraint on annihilating dark matter. For example, using the radio observational data obtained in \cite{Davies} (at 408 MHz from the inner 4 arcsecond cone around Sgr A*) can give a very strong constraint on the dark matter annihilation cross section \cite{Regis,Bertone}. For $m=40$ GeV, the annihilation cross section of the $b\bar{b}$ channel is $<\sigma v> \le 10^{-27}$ cm$^3$ s$^{-1}$ \cite{Bertone,Cholis}. However, the analyses of the data from this very small region have large uncertainties, including the uncertainties of the magnetic field strength and the complicated process of the electron and positron diffusion near the Milky Way center. Moreover, as pointed out in \cite{Cholis}, the effect of inverse Compton scattering might significantly affect the constraints obtained. Some other radio observations also provide good constraints on annihilating dark matter \cite{Borriello,Hooper,Storm,Laha,Wechakama}. However, due to the observational limitations and uncertainties, these constraints are generally less stringent.
 
Besides our Galaxy, M31 is also a good candidate because it is a nearby and well-studied galaxy. Recent radio observations of M31 constrain the dark matter mass $m \ge 100$ GeV and $m \ge 55$ GeV annihilating via $b\bar{b}$ and $\tau^+\tau^-$ channels respectively for $<\sigma v>=3 \times 10^{-26}$ cm$^3$ s$^{-1}$ \cite{Egorov}. However, since the central magnetic field and the dark matter density are poorly constrained for that small observed region ($\approx 1$ kpc), the results have large uncertainties. In this article, we revisit the constraints on annihilating dark matter by using the radio data from \cite{Giebubel}, which originate from a larger region ($\approx 17.5$ kpc) of M31. Also, we model the dark matter density profile of M31 by using the latest data from M31 rotation curve. We mainly focus on six different standard model annihilation channels ($e^+e^-, \mu^+\mu^-, \tau^+\tau^-, u\bar{u}, b\bar{b}$ and $W^+W^-$).

\section{Radio observations of M31}
The group in \cite{Giebubel} uses the Westerbork Synthesis Radio Telescope (WSRT) to observe M31 in the frequency range $\nu=310-376$ MHz. After analyzing the radio data, a uniformly weighted average of the final total power image is obtained for a region of 17.4 kpc. The total flux density $F=(4 \pi D^2)^{-1}dW/d\nu$ integrated over the radius interval $R=0-17.4$ kpc is $F=10.6 \pm 0.7$ Jy, where $D$ is the distance of M31. It is equivalent to the total energy flux $S=\nu F \le 4 \times 10^{-14}$ erg cm$^{-2}$ s$^{-1}$ (2$\sigma$ upper limit). If we assume that all the radio radiation originates from the synchrotron radiation of the electron and positron pairs produced by dark matter annihilation, the above upper limit of the total energy flux can be used to constrain the cross section of dark matter annihilation (no spectral index has to be assumed). From the analyses in \cite{Egorov}, synchrotron radiation dominates the cooling rate of the electron and positron pairs. Therefore, we neglect the effect of the inverse Compton scattering. Also, the cooling processes by other mechanisms such as bremsstrahlung, ionization, scattering, advection loss and re-acceleration are negligible. These processes just contribute 1\% of the total cooling rate. Furthermore, the diffusion time scale of the electron and positron pairs is much longer than the cooling time scale. For a 1 GeV electron, the diffusion and cooling time scales are $t_D \sim R^2/D_0 \sim 10^{17}$ s and $t_c \sim 1/b \sim 10^{16}$ s respectively \cite{Colafrancesco}, where $D_0 \sim 10^{28}$ cm$^2$ s$^{-1}$ is the diffusion coefficient of M31 \cite{Berkhuijsen} and $b \sim 10^{16}$ s$^{-1}$ is the cooling rate. Therefore, the diffusion term can be neglected and the injected spectrum of the electron and positron pairs is proportional to the source spectrum \cite{Storm}.

Since the diffusion process is not important and the radio emissivity is mainly determined by the peak radio frequency (monochromatic approximation), the total synchrotron radiation energy flux of the electron and positron pairs produced by dark matter annihilation is given by \cite{Bertone,Profumo}:
\begin{equation}
S \approx \frac{1}{4 \pi D^2} \left[ \frac{9 \sqrt{3}<\sigma v>}{2m^2} \int_0^R 4 \pi r^2 \rho_{DM}^2EY(E)dr \right],
\end{equation}
where $D=785 \pm 25$ kpc \cite{Egorov}, $\rho_{DM}$ is the dark matter density profile of M31, $E=0.43(\nu/{\rm GHz})^{1/2}(B/{\rm mG})^{-1/2}$ GeV, and $Y(E)=\int_E^m(dN_e/dE')dE'$. Here, $B$ is the magnetic field strength in M31 and $dN_e/dE'$ is the electron or positron spectrum of dark matter annihilation. The electron or positron spectrum for each annihilation channel can be obtained in \cite{Cirelli}. The magnetic field strength in M31 is quite uniform for $r=6-14$ kpc, which is about $4.6-5.2$ $\mu$G \cite{Fletcher}. For the outer region, the magnetic field is about 4 $\mu$G with a weak radial dependence \cite{Granados}. In the following analysis, we follow \cite{Giebubel} to use $B=5 \pm 1$ $\mu$G for M31. Therefore, the peak energy used in Eq.~(1) is $E=3.1-4.2$ GeV. Since the magnetic field is much stronger near the center of M31, the larger value of $B$ would give a smaller value of $E$ and a larger value of $Y(E)$. However, it is not easy to determine the magnetic field strength profile precisely near the M31 center. Studies in \cite{Egorov,Giebubel2} point out that the magnetic field structure of the central region in M31 is very complicated. The magnetic field strength can vary from 10 $\mu$G to 50 $\mu$G in different regions \cite{Egorov,Giebubel2}. In fact, the dependence of the magnetic strength in Eq.~(1) is quite weak. A factor of 10 larger in $B$ would just give less than a few percent larger in $S$. Therefore, we use a constant profile of $B=5 \pm 1$ $\mu$B to model the magnetic field strength of M31. This would underestimate the total radio flux $S$ calculated by Eq.~(1). Nevertheless, the underestimated value of $B$ can give a conservative lower limit of $S$ for dark matter annihilation.

For the dark matter density profile, recent analysis of the M31 rotation curve gives a robust set of parameters with small errors. Sofue (2015) \cite{Sofue} shows that the NFW profile \cite{Navarro} is likely to be a realistic approximation to model the dark matter density profile of M31:
\begin{equation}
\rho_{DM}=\frac{\rho_sr_s^3}{r(r_s+r)^2},
\end{equation}
where $\rho_s=(2.23 \pm 0.24) \times 10^{-3}M_{\odot}$ pc$^{-3}$ and $r_s=34.6 \pm 2.1$ kpc \cite{Sofue}. Besides the NFW profile, we also examine two other popular profiles, the Burkert profile $\rho_{DM}=\rho_sr_s^3[(r_s+r)(r_s^2+r^2)]^{-1}$ and the Einasto profile $\rho_{DM}=\rho_s\exp\{-17.668[(r/r_s)^{1/6}-1] \}$ \cite{Tamm}. The corresponding parameters for the Burkert profile and the Einasto profile are $(\rho_s,r_s)=(3.68 \pm 0.40 \times 10^{-2}M_{\odot}~{\rm pc}^{-3},9.06 \pm 0.53~{\rm kpc})$ and $(\rho_s,r_s)=(8.12 \pm 0.16 \times 10^{-6}M_{\odot}~{\rm pc}^{-3},178\pm 18~{\rm kpc})$ respectively \cite{Tamm}.

By putting the above different density profiles into Eq.~(1) and using the lower limits of $\rho_s$ and $r_s$ and the upper limit of $D$, we can get an analytic expression for the lower limit of $S$:
\begin{equation}
S \ge S_0 \left( \frac{<\sigma v>}{2.2 \times 10^{-26}~\rm cm^3~s^{-1}} \right) \left(\frac{m}{\rm GeV}\right)^{-2} \left( \frac{E}{\rm GeV} \right)Y(E),
\end{equation}
where $S_0=1.34 \times 10^{-10}$ erg cm$^{-2}$ s$^{-1}$ for the NFW profile, $S_0=1.12 \times 10^{-10}$ erg cm$^{-2}$ s$^{-1}$ for the Burkert profile and $S_0=2.89 \times 10^{-10}$ erg cm$^{-2}$ s$^{-1}$ for the Einasto profile. By using the $Y(E)$ values shown in Fig.~1 and assuming the canonical thermal relic annihilation cross section, we can get the lower limit of $S$ for different annihilation channels and different density profiles (see Figs.~2-4). For the most popular $b\bar{b}$ annihilation channel, the lower bound of $m$ is 120 GeV for the NFW profile. This is a bit tighter than the constraint obtained in the gamma-ray observations of the MW dSphs ($m \ge 100$ GeV) \cite{Ackermann} and the previous radio observation of M31 ($m \ge 90$ GeV for the $b\bar{b}$ channel with $<\sigma v>=2.2 \times 10^{-26}$ cm$^3$ s$^{-1}$) \cite{Egorov}. If we use the Burkert profile, the lower limits would be smaller by about 10-20\%. Although the Burkert profile is not a robust profile for large galaxy such as M31, we still consider these lower limits the most conservative limits of our analyses. In table 1, we summarize the conservative lower limits of the dark matter mass for $<\sigma v>=2.2 \times 10^{-26}$ cm$^3$ s$^{-1}$.

\begin{table}
\caption{Lower limits of the dark matter mass for different dark matter density profiles. Here, we assume $<\sigma v>=2.2 \times 10^{-26}$ cm$^3$ s$^{-1}$.}
 \label{table1}
 \begin{tabular}{@{}lccc}
  \hline
   & NFW & Burkert & Einasto \\
  \hline
  $e^+e^-$ & 190 GeV & 170 GeV & 270 GeV \\
  $\mu^+\mu^-$ & 100 GeV & 90 GeV & 140 GeV \\
  $\tau^+\tau^-$ & 90 GeV & 80 GeV & 140 GeV \\
  $u\bar{u}$ & 110 GeV & 90 GeV & 230 GeV \\
  $b\bar{b}$ & 120 GeV & 90 GeV & 250 GeV \\
  $W^+W^-$ & 90 GeV & 90 GeV & 200 GeV \\
  \hline
 \end{tabular}
\end{table}

If we assume that the annihilation cross section is a free parameter, we can compare our results with the constraints obtained in the recent gamma-ray observations of the MW dSphs \cite{Ackermann} and the Milky Way center \cite{Calore2,Abazajian2}. In Fig.~5, our upper limits are 1-2 orders of magnitude tighter than the 95\% C.L. upper limits obtained in the MW dSphs for the $e^+e^-$ and $\mu^+\mu^-$ channels for $m=10-1000$ GeV. It is because a large amount of high-energy electron-positron pairs is produced in these two channels. The predicted synchrotron radiation signals are very large and hence the constraints are more stringent. For the other channels, our results are generally tighter when $m$ is smaller than $\sim 100$ GeV (except the $u\bar{u}$ channel). In Fig.~6, we compare our results with the recent empirical fits of the Galactic GeV excess obtained in \cite{Calore2,Abazajian2}. We can see that our constraints rule out the best models of dark matter interpretation of the GeV excess for the $b\bar{b}$, $\mu^+\mu^-$ and $\tau^+\tau^-$ channels (by at least 2$\sigma$). Nevertheless, the parameters of the $u\bar{u}$ channel can still satisfy our constraints marginally. Since the studies in \cite{Ackermann,Calore2} assume the NFW profile to calculate the limits, we also use the NFW dark matter profile to do the analysis.

If we allow mixed annihilations, Calore et al. (2015) \cite{Calore2} predict that the ratio $b\bar{b}:c\bar{c}:\tau^+\tau^-=0.87:0.08:0.05$ ($bc\tau$ model) would be the best to account for the Milky Way GeV gamma-ray excess. A good fit can also be obtained if the annihilation products are $\mu^+\mu^-$ and $\tau^+\tau^-$ ($\mu\tau$ model) for $m \sim 50$ GeV with the branching ratio of $\mu^+\mu^-$ $\ge 0.6$ \cite{Calore2}. If we assume $m=50$ GeV, our results rule out the $bc\tau$ model and $\mu\tau$ model by the 1$\sigma$ and 2$\sigma$ radio upper limit respectively. Therefore, based on our analyses, our new constraints do not favor the dark matter interpretation of the Milky Way GeV gamma-ray excess. Our results support the conclusion drawn from the Fermi-LAT gamma-ray observations of the MW dSphs \cite{Ackermann}.

\begin{figure}
\vskip 10mm
 \includegraphics[width=82mm]{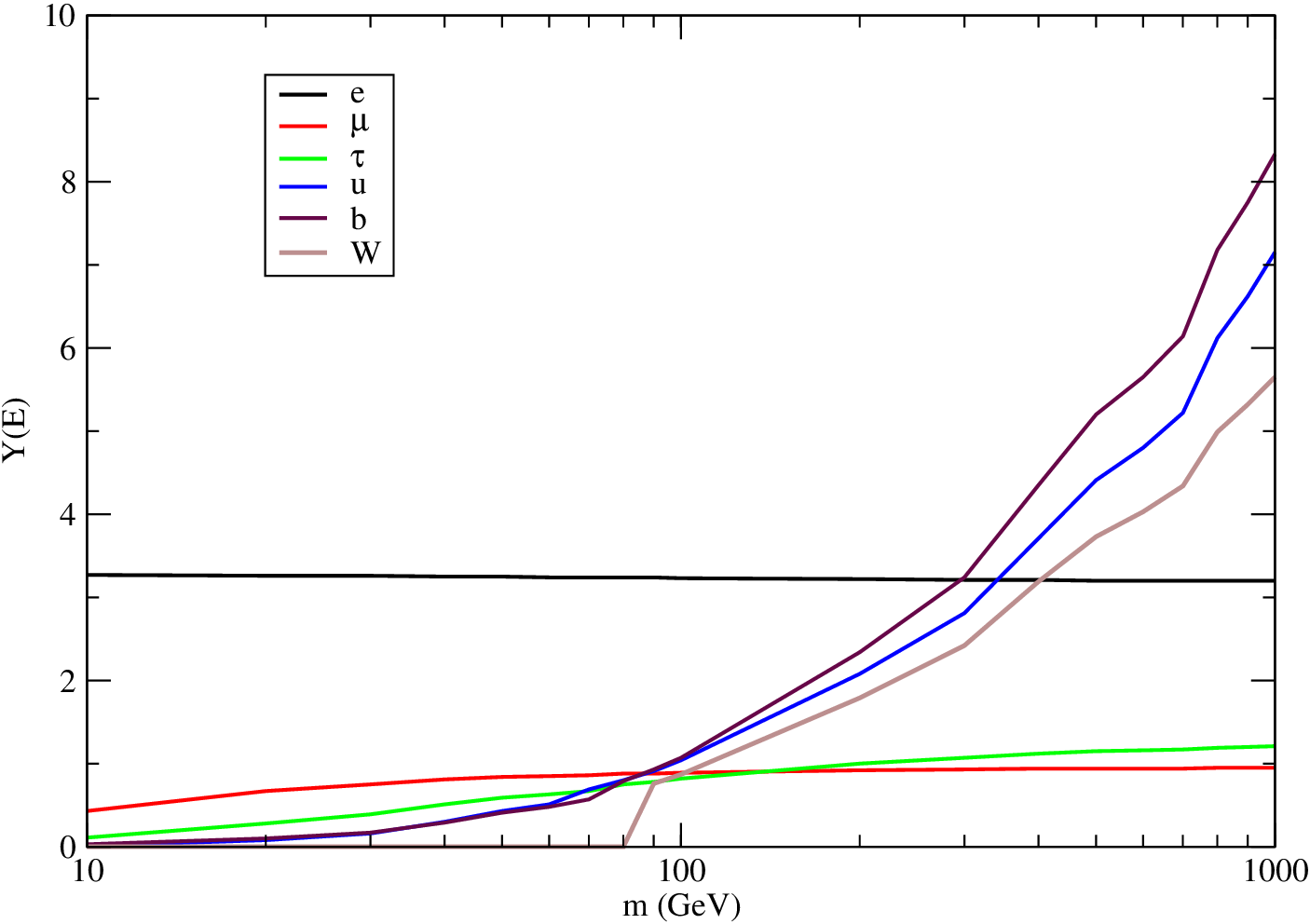}
 \caption{The graph of $Y(E)$ versus $m$ for the six annihilation channels. Here, we use $E=3.5$ GeV.}
\vskip 10mm
\end{figure}

\begin{figure}
\vskip 10mm
 \includegraphics[width=82mm]{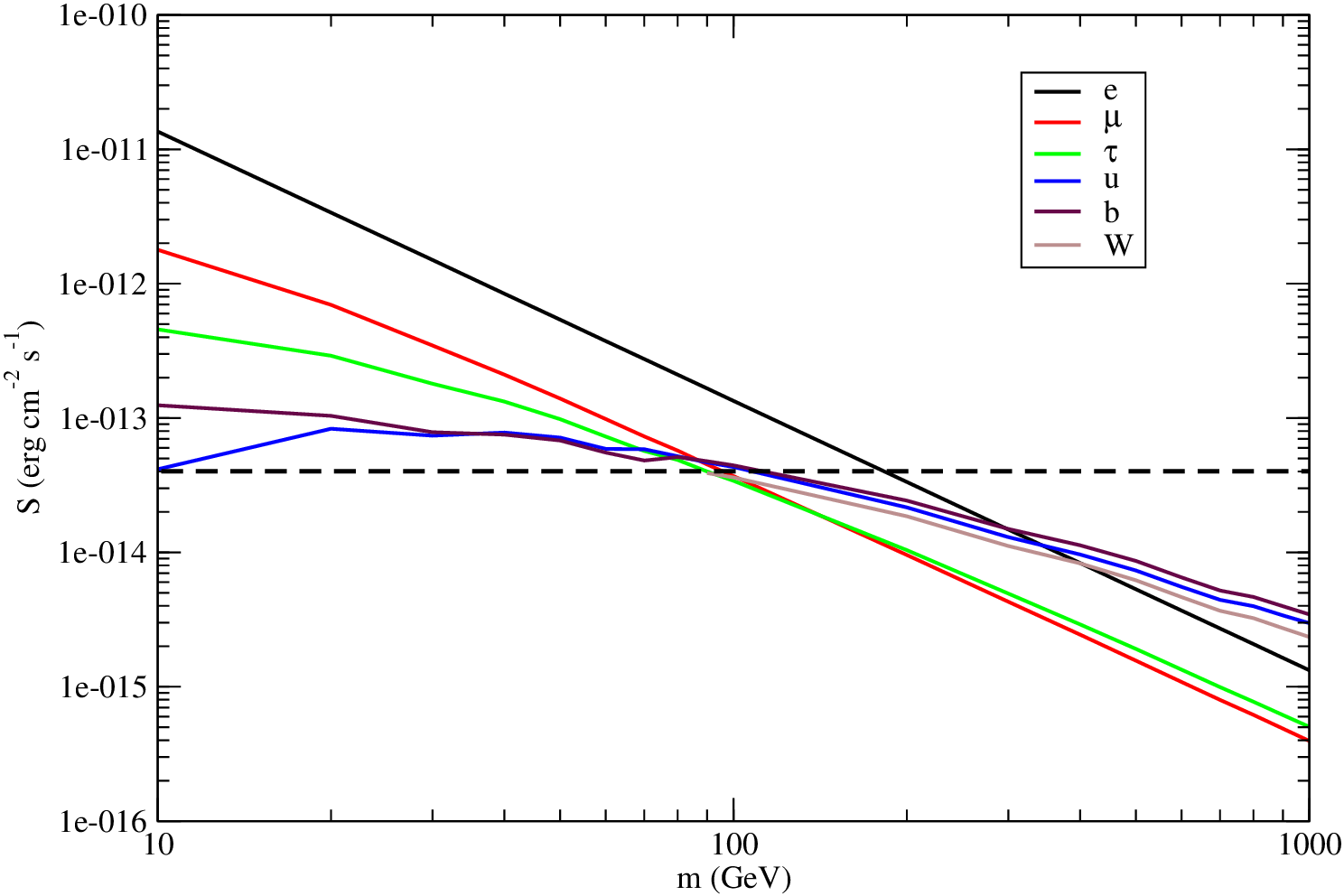}
 \caption{The minimum values of $S$ for $<\sigma v>=2.2 \times 10^{-26}$ cm$^3$ s$^{-1}$ with the NFW density profile. The dashed line is the 2$\sigma$ upper limit of the radio observations \cite{Giebubel}.}
\vskip 10mm
\end{figure}

\begin{figure}
\vskip 10mm
 \includegraphics[width=82mm]{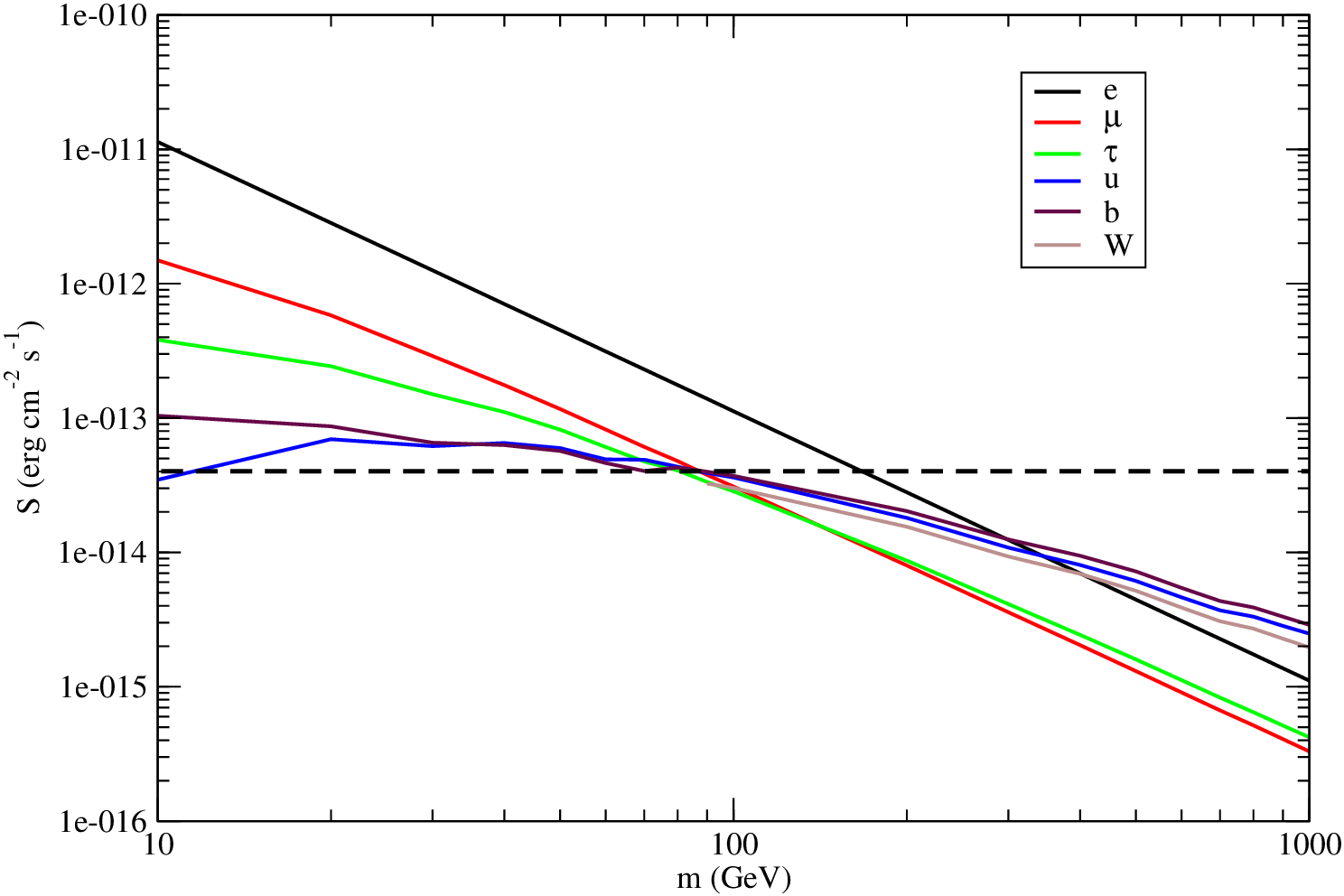}
 \caption{The minimum values of $S$ for $<\sigma v>=2.2 \times 10^{-26}$ cm$^3$ s$^{-1}$ with the Burkert density profile. The dashed line is the 2$\sigma$ upper limit of the radio observations \cite{Giebubel}.}
\vskip 10mm
\end{figure}

\begin{figure}
\vskip 10mm
 \includegraphics[width=82mm]{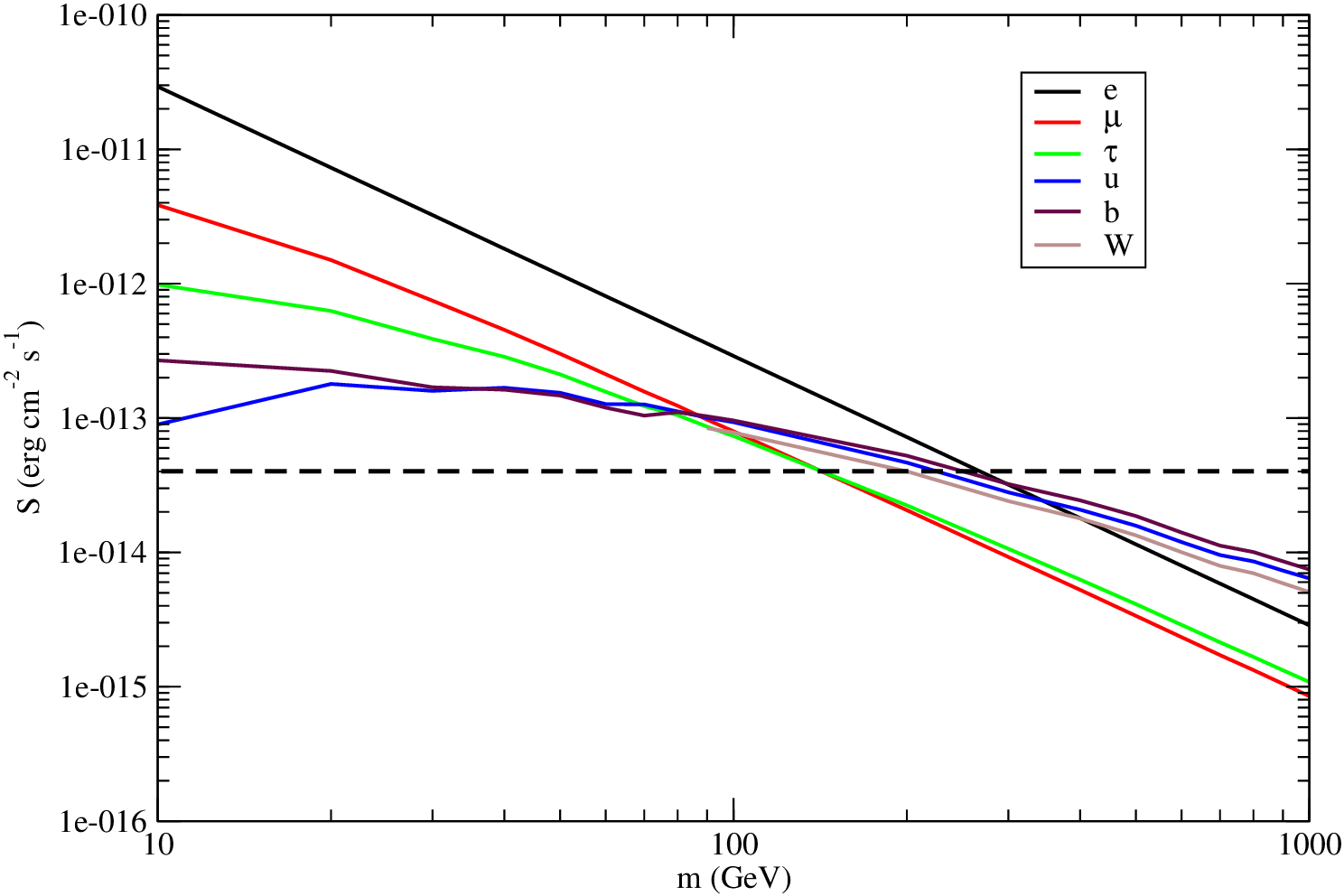}
 \caption{The minimum values of $S$ for $<\sigma v>=2.2 \times 10^{-26}$ cm$^3$ s$^{-1}$ with the Einasto density profile. The dashed line is the 2$\sigma$ upper limit of the radio observations \cite{Giebubel}.}
\vskip 10mm
\end{figure}

\begin{figure}
\vskip 10mm
 \includegraphics[width=82mm]{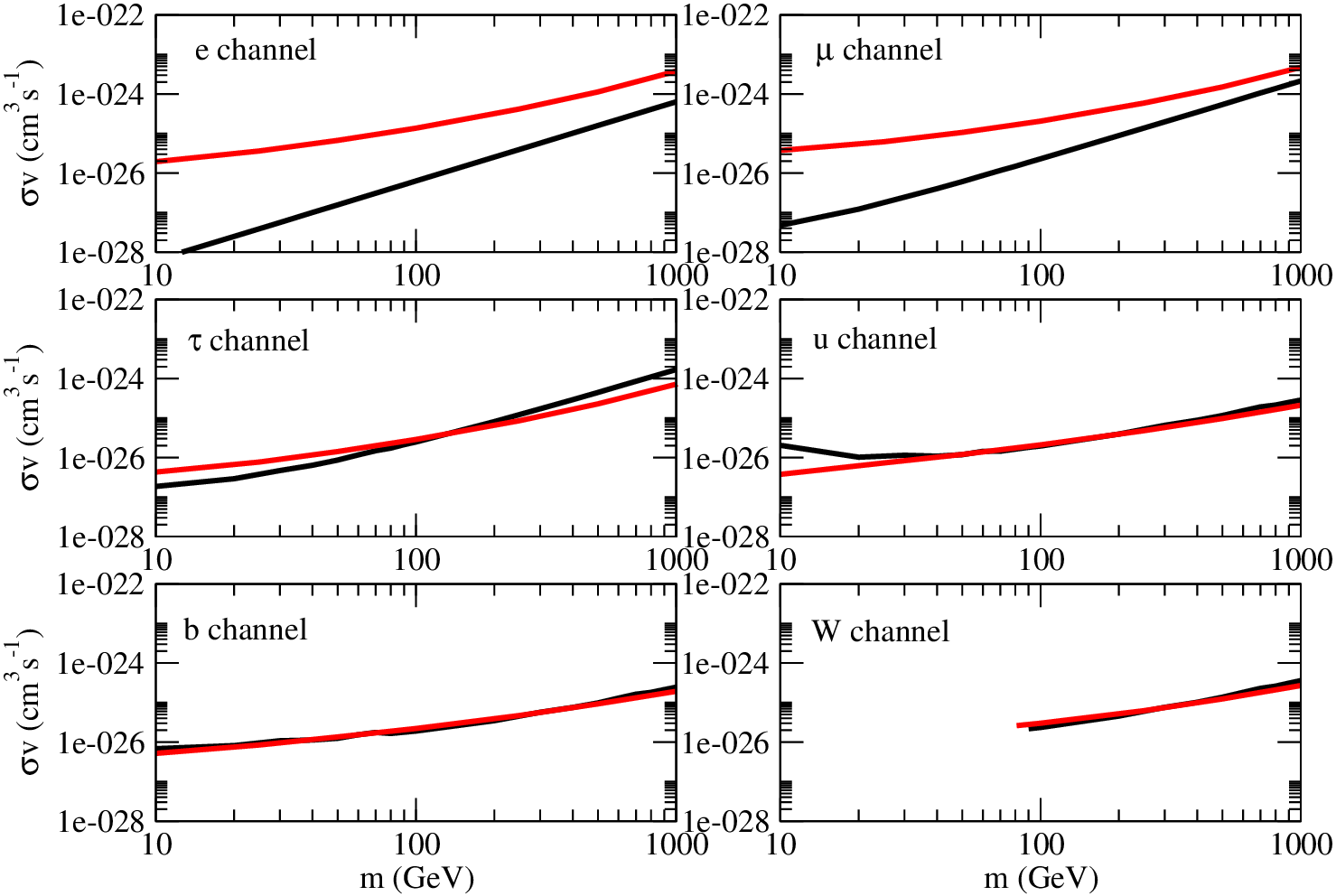}
 \caption{The upper limits of the annihilation cross sections for the six annihilation channels (assumed NFW profile) (black: our results; red: gamma-ray observations of Milky Way dwarf spheroidal satellite galaxies \cite{Ackermann}.)}
\vskip 10mm
\end{figure}

\begin{figure}
\vskip 10mm
 \includegraphics[width=82mm]{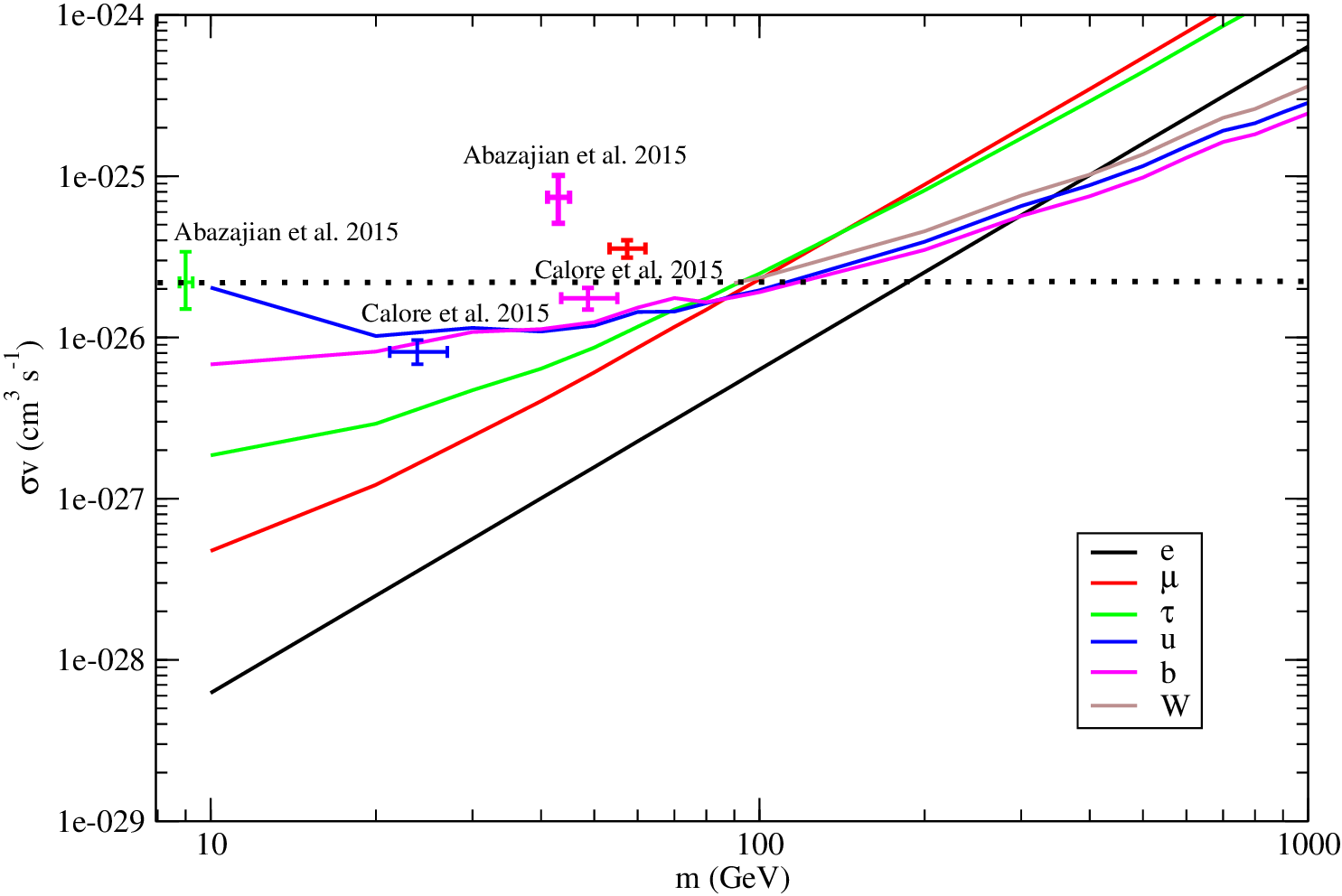}
 \caption{The upper limits of the annihilation cross sections for the six annihilation channels (assumed NFW profile). The data points with 1$\sigma$ error bars are the results obtained in \cite{Calore2,Abazajian2} for the dark matter interpretation of the GeV excess. The dotted line is the canonical thermal relic cross section.}
\vskip 10mm
\end{figure}

\section{Discussion}
In this article, we revisit the radio constraints of annihilating dark matter by using the M31 radio data in \cite{Giebubel}. Our results generally give more stringent constraints on annihilation cross sections for the six standard model annihilation channels. Here, the uncertainties of the parameters involved in the analyses, such as the scale density $\rho_s$, the scale radius $r_s$, magnetic field strength $B$ and the observed radio flux $S$, are relatively small compared with previous studies. Since the magnetic field strength is greater in the center of M31, the radio flux emitted by the electron-positron pairs produced from dark matter annihilation should be larger. Therefore, our assumption of the constant magnetic field strength gives a conservative lower limits of the radio emission. We find that all of the conservative lower limits are larger than the 2$\sigma$ upper limits of the observed flux for $m \ge 80$ GeV if we assume a canonical thermal relic cross section. It is consistent with the results obtained by gamma-ray observations of the MW dSphs \cite{Ackermann}. Therefore, our result provides an independent support of the recent analysis that most of the standard $10-100$ GeV dark matter annihilation models should be ruled out.

If we release the annihilation cross section to be a free parameter, we also obtain constraints of the annihilation cross sections for the six channels. Generally speaking, our constraints are more stringent than that obtained in \cite{Ackermann}, especially for the $e^+e^-$ channel, $\mu^+\mu^-$ channel, $\tau^+\tau^-$ channel for $m \le 140$ GeV, $b\bar{b}$ channel for $m \approx 40-60$ GeV and $W^+W^-$ for $m \le 300$ GeV. These constraints are useful to examine the most popular dark matter interpretation of the GeV excess \cite{Calore2,Abazajian2}, such as the $b\bar{b}$ and $\tau^+\tau^-$ channels. Based on our analyses, most of the models are ruled out except the $u\bar{u}$ model. Our analyses also show that the $bc\tau$ mixed annihilation model and the $\mu\tau$ mixed annihilation model exceed the 1$\sigma$ and 2$\sigma$ upper limit of the radio flux respectively. If we can get some better radio data of M31 or precisely determine the magnetic field profile in the future, more stringent constraints can be obtained. 

\section{acknowledgements}
This work is partially supported by a grant from the Education University of Hong Kong (Project No.:RG57/2015-2016R).

\end{document}